\documentclass[aps,prb,twocolumn,superscriptaddress,showpacs,longbibliography]{revtex4-1}

\usepackage{hyperref}
\usepackage[dvips]{graphicx}
\usepackage{amssymb,amsmath,amsfonts}
\usepackage[T1]{fontenc}
\usepackage{float}
\usepackage{color,soul}
\usepackage{booktabs}
\usepackage{color}
\newcommand{\bra}[1]{\langle #1|}
\newcommand{\ket}[1]{|#1\rangle}
\newcommand{\unit}{$(\hbar/e)(\Omega\times $cm$)^{-1}$}

\begin{document}


\title{Spin Hall effect in prototype Rashba ferroelectrics GeTe and SnTe}


\author{Haihang Wang}
\affiliation{Department of Physics, University of North Texas, Denton, TX 76203, USA}

\author{Priya Gopal}
\affiliation{Department of Physics, University of North Texas, Denton, TX 76203, USA}

\author{Silvia Picozzi}
\affiliation{Consiglio Nazionale delle Ricerche, Istituto SPIN, UOS L'Aquila, Sede di lavoro CNR-SPIN c/o Univ. "G. D'Annunzio", 66100 Chieti, Italy}

\author{Stefano Curtarolo}
\email{stefano@duke.edu}
\affiliation{Center for Materials Genomics, Duke University, Durham, NC 27708, USA}
\affiliation{Materials Science, Electrical Engineering, Physics and Chemistry, Duke University, Durham, NC 27708, USA}

\author{Marco \surname{Buongiorno Nardelli}}
\affiliation{Department of Physics, University of North Texas, Denton, TX 76203, USA}
\affiliation{Center for Materials Genomics, Duke University, Durham, NC 27708, USA}

\author{Jagoda S\l awi\'{n}ska}
\email{jagoda.slawinska@unt.edu}
\affiliation{Department of Physics, University of North Texas, Denton, TX 76203, USA}

\vspace{.15in}

\date{\today}

\begin{abstract}
Ferroelectric Rashba semiconductors (FERSC) have recently emerged as a promising class of spintronics materials. The peculiar coupling between spin and polar degrees of freedom responsible for several exceptional properties, including ferroelectric switching of Rashba spin texture, suggests that the electron's spin could be controlled by using only electric fields. In this regard, recent experimental studies revealing charge-to-spin interconversion phenomena in two prototypical FERSC, GeTe and SnTe, appear extremely relevant. Here, by employing density functional theory calculations, we investigate spin Hall effect (SHE) in these materials and show that it can be large either in ferroelectric or paraelectric structure. We further explore the compatibility between \textit{doping} required for the practical realization of SHE in semiconductors and \textit{polar distortions} which determine Rashba-related phenomena in FERSC, but which could be suppressed by free charge carriers. Based on the analysis of the lone pairs which drive ferroelectricity in these materials, we have found that the polar displacements in GeTe can be sustained up to a critical hole concentration of over $\sim 10^{21}$/cm$^{3}$, while the tiny distortions in SnTe vanish at a minimal level of doping. Finally, we have estimated spin Hall angles for doped structures and demonstrated that the spin Hall effect could be indeed achieved in a polar phase. We believe that the confirmation of spin Hall effect, Rashba spin textures and ferroelectricity coexisting in one material will be helpful for design of novel multifunctional spintronics devices operating without magnetic fields.

\end{abstract}

\maketitle

\section{Introduction}

Employing electron's spin for information processing is one of the main goals of semiconductor spintronics.\cite{spintronics, das} The electric and nonvolatile control of spins in recently discovered ferroelectric Rashba semiconductors (FERSC) holds a promise to combine storage, memory and computing functionalities.\cite{fersc, silvia} The exceptional properties of these novel materials arise from the unique property that the spin degrees of freedom are coupled to the electric polarization, whereby the polar axis intrinsically breaks the inversion symmetry (IS). The giant Rashba spin-splitting of the bulk electronic states can be then switched with the ferroelectric orientation and controlled by an electric field. On the other hand, the large spin-orbit coupling (SOC) required in these materials could also give rise to charge-to-spin interconversion phenomena, such as spin Hall effect (SHE),\cite{sinova, she1, kato, sinova_she} Edelstein-Rashba effect\cite{rashba-edelstein} or spin galvanic effect\cite{galvanic} allowing the electric control of spin currents.\cite{awschalom} In this paper, we have theoretically studied the intrinsic spin Hall effect in two prototype FERSC compounds, GeTe and SnTe, and explored its compatibility with the polar distortion, a fundamental property in these materials.

The ferroelectric XTe (X=Ge,Sn) monochalcogenides share rhombohedrally distorted rocksalt structure (space group R3m). Both possess spontaneous electric polarization $\vec{P}$ parallel to [111] direction, but the high Curie temperature of GeTe ($T_C$ = 700 K)\cite{rabe, helmut} allowed in-depth experimental studies of the ferroelectric phase which proved the polarization switching,\cite{switching_gete} electronic structures\cite{liebmann, minar1, minar2} and reversible Rashba spin textures.\cite{gete_nano} Importantly, spin-to-charge conversion has been recently measured via spin pumping experiments in Fe/GeTe bilayers demonstrating high potential of such interfaces for multifunctional spintronics applications.\cite{fert, fe-gete} On the other hand, ferroelectric phase of SnTe ($T_C$ = 100 K)\cite{snte_temp} has been known mainly from the theoretical side.\cite{plekhanov} Although the intermediate topological phases predicted at the phase transition still await an experimental confirmation, even the room temperature cubic crystal is intriguing, as it represents a prototypical example of a topological crystalline insulator.\cite{tci_snte} Finally, recent spin pumping experiments performed for the high-symmetry paraelectric bulk revealed strong spin Hall effect whose origin remains elusive.\cite{ohya_snte}

Our calculations based on density functional theory (DFT) allowed us to quantitatively estimate spin Hall conductivities (SHC) for low- and high-symmetry structures of both materials. First, we have unveiled that the ferroelectric phase could indeed enhance the spin Hall effect as compared with the paraelectric structure. We have interpreted this effect in terms of additional contributions to spin Berry curvature, originating from spin-splitted electronic states in the polar phase. Second, since the realization of SHE in semiconductors requires doping, we have studied the evolution of polar distortion with respect to the charge carrier concentration and estimated critical levels of doping that can sustain the low-symmetry phase. Finally, we have calculated spin Hall angles for doped structures and explored their potential for practical realization of spin Hall effect and electric control of Rashba effect at the same time.

The paper is organized as follows: Section II summarizes the details of DFT and spin Hall conductivity calculations. In Section III, we briefly describe electronic structures of GeTe/SnTe and discuss spin Hall effect in ferroelectric and paraelectric configurations. Section IV is focused on the evolution of polar distortions in the presence of charge carriers. Section V reports spin Hall angles calculated for doped FERSC. The conclusions and perspectives are given in Section VI.

\section{Computational details}

\begin{figure}
\includegraphics[width=\columnwidth]{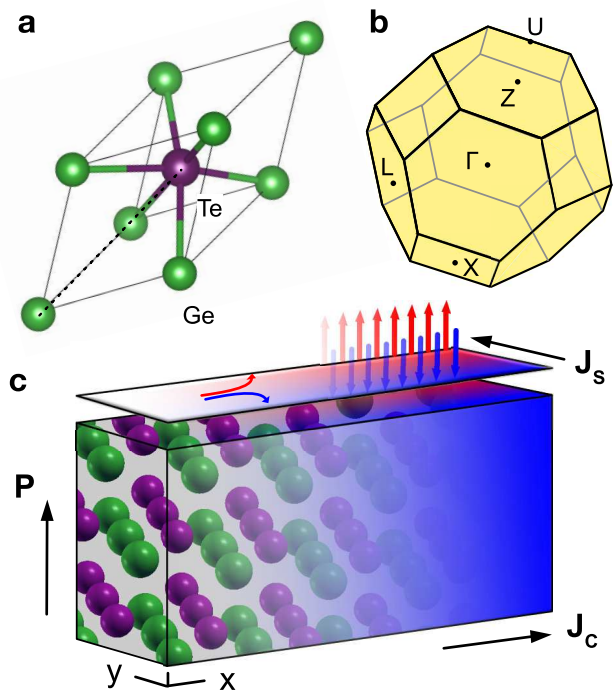}
\caption{\label{struct}
    Structure of prototype Rashba ferroelectrics. (a) Crystal structure of distorted GeTe with the polar axis along [111] direction. (b) BZ corresponding to the rhombohedral unit cell shown in (a). The labels denote high-symmetry points used to evaluate band structures. (c) Schematic view of spin Hall effect in the configuration $\sigma^{z}_{xy}$. Directions of charge and spin currents are marked by the arrows. Red and blue shades denote spin polarization parallel and antiparallel to electric polarization, respectively.
}
\end{figure}

First-principles calculations based on density functional theory (DFT) were performed using \textsc{Quantum Espresso} package\cite{qe,qe1} interfaced with the \textsc{Aflow$\pi$} infrastructure.\cite{aflowpi} We treated the exchange and correlation interaction within the generalized gradient approximation (GGA),\cite{pbe} and the ion-electron interaction  with the projector augmented-wave fully-relativistic pseudopotentials\cite{kresse-joubert} from the pslibrary database.\cite{pslibrary} The electron wave functions were expanded in a plane wave basis set with the cutoff of 85 Ry. The description of electronic structures was corrected by using a novel pseudo-hybrid Hubbard self-consistent approach ACBN0.\cite{acbn0} The rhombohedral unit cells schematically shown in Fig. \ref{struct} were fully relaxed until the forces on each atom became smaller than $10^{-3}$ Ry/bohr; the optimized lattice constants of the ground-state ferroelectric GeTe (SnTe) are equal to 4.37 \AA\, (4.57 \AA), respectively. The Brillouin zone sampling at the level of DFT was performed following the Monkhorst-Pack scheme using a $16\times16\times16$ k-points grid. The electron/hole doping was simulated by adding extra charge to the system with the compensating amount of opposite charge in the background.

The intrinsic spin Hall conductivities were calculated as implemented in the \textsc{Paoflow} code\cite{paoflow} following the Kubo's formula:\cite{kubo, gradhand}

\begin{eqnarray}
    \sigma^{k}_{ij} = \frac{e^2}{\hbar}\sum_{\vec{k}}\sum_{n}f_n(\vec{k})\Omega^{k}_{n,ij}(\vec{k})
    \nonumber
\end{eqnarray}
\begin{eqnarray}
    \Omega^{k}_{n,ij} (\vec{k}) = \sum_{m\neq n}\frac{2\textrm{Im}\bra{\psi_{n, \vec{k}}}j^{k}_{i}\ket{\psi_{m, \vec{k}}}\bra{\psi_{m, \vec{k}}}v_{j}\ket{\psi_{n, \vec{k}}}}{(E_n-E_m)^2}\nonumber
\end{eqnarray}

\noindent where $\Omega^{k}_{n,ij}$ is the spin Berry curvature of $n$th band assuming spin polarization along $k$ and spin (charge) current along $i$ ($j$), $f_n(\vec{k})$ is the Fermi-Dirac distribution function and $j^{k}_{i} = \{s_{k}, v_{i}\}$ is the spin current operator, where $s$ and $v$ stand for spin and velocity operators, respectively. In order to properly resolve the rapid variation of spin Berry curvatures, we have interpolated the $k$-points mesh to $96\times96\times96$ using the adaptive smearing in the final spin Hall conductivity calculations.


\section{Electronic properties and spin Hall effect}
\begin{figure*}[ht!]
    \includegraphics[width=\textwidth]{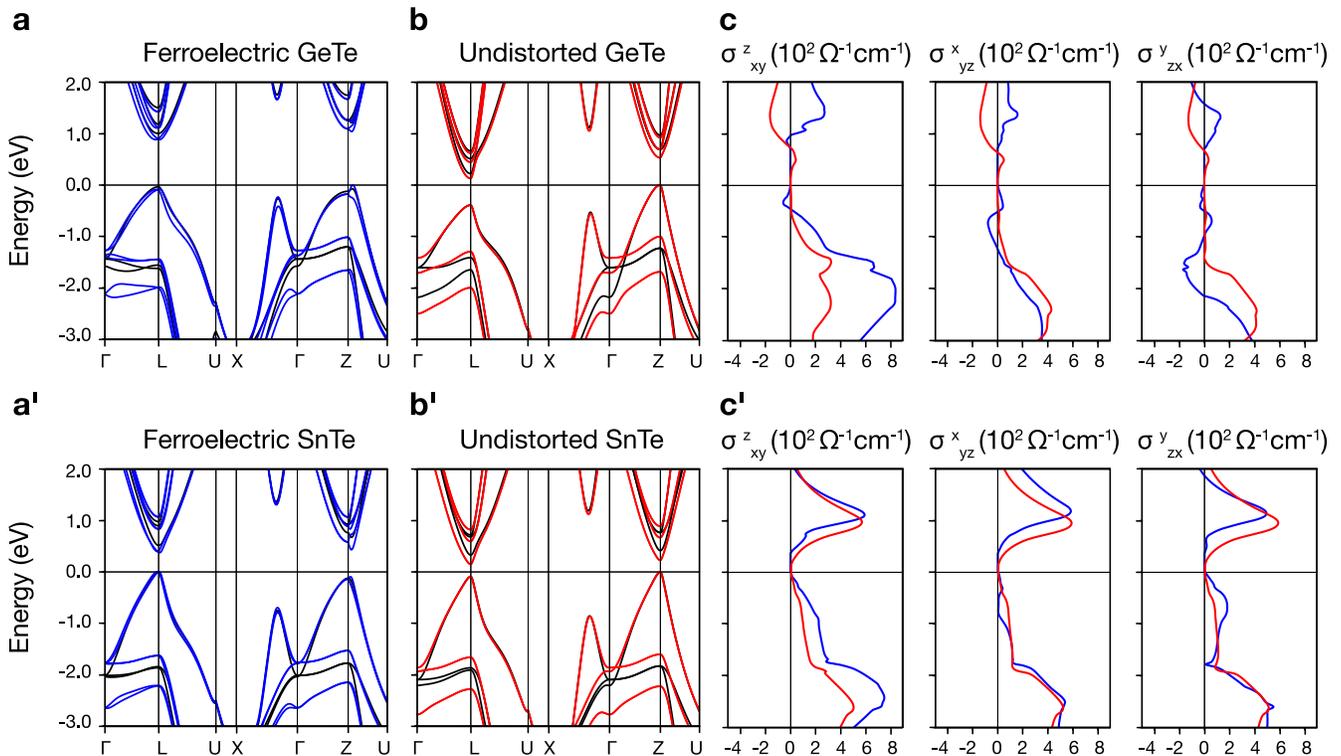}
    \caption{\label{bands}
    Electronic properties and spin Hall conductivities of GeTe and SnTe. Panels (a) and (b) show fully relativistic band structures of distorted (blue) and centrosymmetric (red) GeTe, respectively. In order to simplify analysis, we have used same rhombohedral unit cell in both (a) and (b). Black lines denote results of scalar-relativistic calculations. (c) Different components of spin Hall conductivity tensor as a function of chemical potential. Blue and red lines correspond to distorted and undistorted structures in accordance with band structures in (a-b). We note that there are only four non-zero independent elements of $\sigma^{k}_{ij}$; those selected above could correspond to real SHE geometries. (a'-c') Same as (a-c) for SnTe. The Fermi energy is set at the VBM in all panels. We also note that the CBM does not lie along any of the high-symmetry lines.
    }
\end{figure*}

Spin Hall materials are usually found either among elemental metals with strong SOC\cite{pt, au_pd, ta} or novel topological phases, such as Weyl and Dirac nodal line semimetals where spin-orbit protected gaps between the electronic bands are crossed by the Fermi level.\cite{taas, felser, iro2} It has also been known that a giant Rashba effect (RE) could induce large SHC as long as the Fermi level can be tuned and brought between lower and upper Rashba bands.\cite{xiao_rashba} In a polar material, the inversion symmetry breaking gives rise to giant Rashba spin-orbit derived spin splittings of the electronic states. Therefore, we expect that the ferroelectric structure could, in principle, enhance the spin Hall conductivity over the paraelectric phase at specific values of the chemical potential.

Before we start a more detailed analysis of the relationship between electronic structures and spin Hall conductivities in polar and nonpolar phases, let us briefly comment on the accuracy of our calculations, essential for realistic evaluation of the SHC. Figure \ref{bands}(a) shows the band structure calculated for the ferroelectric GeTe (blue line) along the high-symmetry directions defined in Fig. \ref{struct} (b). The ferroelectric distortion of the rhombohedral structure gives rise to inequivalent hexagonal faces in the Brillouin zone (BZ), two of them perpendicular to the polar axis and centered at $Z$ points, and six corresponding to other hexagons centered at $L$, as marked in the scheme. The band structure agrees perfectly with earlier theoretical predictions and experimental results. In particular, our band gap $E_g=0.65$ eV is very close to the experimental one (0.61 eV)\cite{gete_gap} and coincides exactly with the value obtained previously with Heyd-Scuseria-Ernzerhof (HSE) hybrid functional.\cite{fersc} We note that a simple PBE calculation yields an incorrect energy gap of only 0.3 eV. Importantly, the reliable value of the band gap is relevant for the estimation of Rashba spin splittings. They can be easily noticed in Fig. \ref{bands} (a), especially close to the valence band maximum (VBM) and highest conduction band minimum (HBM) along the $ZU$ direction. Their values, 250 meV and 100 meV for valence and conduction bands, respectively, are again in excellent agreement with the HSE calculations and with experimental results reporting giant Rashba effect in bulk electronic states of GeTe.\cite{fersc, minar2, gete_nano}
\vskip 1.0cm

Surprisingly, despite much smaller polar displacement the electronic properties of the ferroelectric SnTe are quite similar to those described above. The strong SOC combined with the small band gap (0.33 eV) induces giant Rashba splittings of the valence and conduction bands along $ZU$ path, estimated to approximately 290 and 180 meV, respectively, again in good agreement with earlier hybrid functional calculations.\cite{plekhanov}

As a next step, we will analyze spin Hall effect in the two compounds. Figure \ref{bands}(c-c') shows spin Hall conductivity vs energy  calculated for three selected components of $\sigma^{k}_{ij}$. We can observe that all SHC curves reveal apparently large spin Hall effect in either ferroelectric (blue) or centrosymmetric (red) configuration. Spin Hall effect in nonpolar structures originates from the significant impact of SOC on the electronic states, as deduced from the comparison between fully-relativistic (red) and scalar-relativistic (black) electronic structures (b-b'). The same contribution from huge bands anticrossings is present in the ferroelectrics, but it is clear that in some BZ regions the magnitudes of SHC are enhanced with respect to the undistorted structures; we attribute it to the Rashba spin-splittings due to the IS breaking.

The most pronounced difference between polar and nonpolar configurations can be found for $\sigma^{z}_{xy}$ component at energies between -2.0 and -3.0 eV. We anticipate, however, that such a high level of doping cannot be achieved in practice. Instead, the features that really contribute to SHE in polar GeTe are small peaks of $\sigma^{z}_{xy}$ and $\sigma^{x}_{yz}$ just below the Fermi level, the former clearly originating from the strong spin-splitting of VBM along $ZU$ direction. Moreover, we emphasize that in practical realizations band structures might change upon charge injection, thus the analysis based on the rigid shift of the chemical potential in Fig.\ref{bands}(c-c') provides only a rough approximation of SHC in doped materials. In order to obtain more realistic predictions, we have performed additional calculations which explicitly takes charge carriers into account.

\section{Doping-induced evolution of polar displacements}
\begin{figure*}[ht!]
    \includegraphics[width=\textwidth]{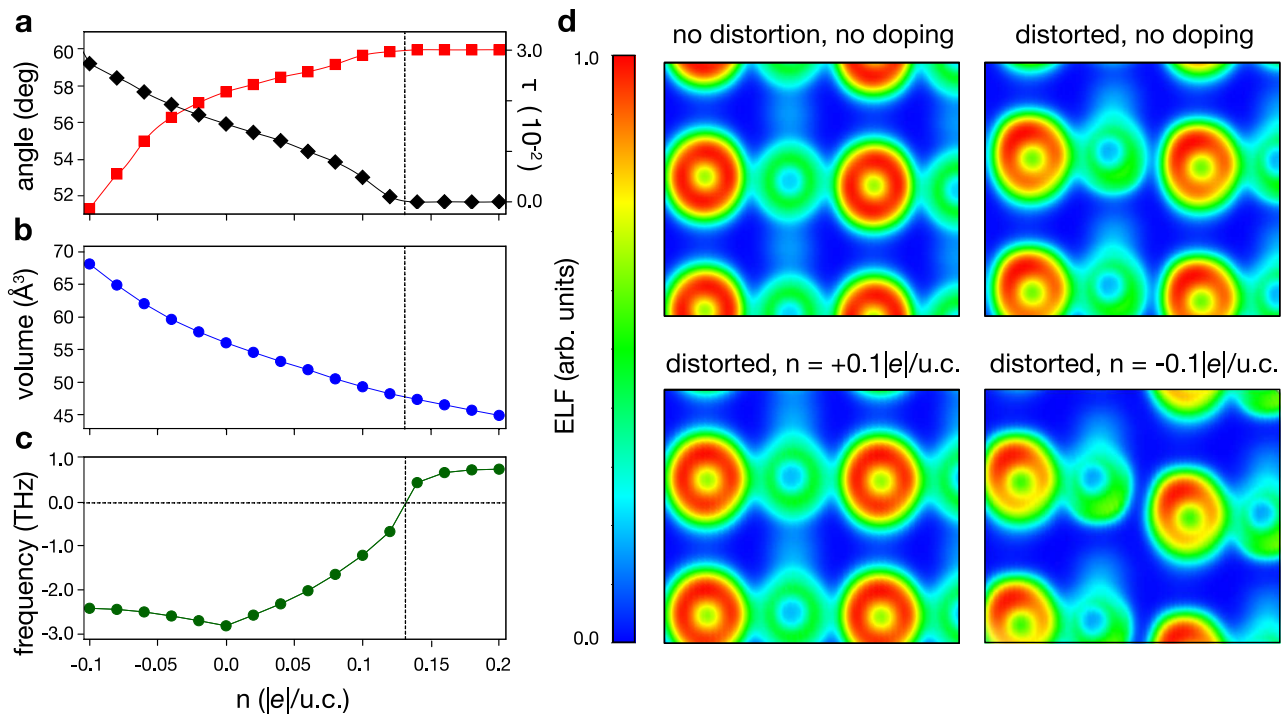}
    \caption{\label{doping}
   Evolution of crystal distortions with varying doping concentration in GeTe. Panel (a) shows the relative displacement $\tau$ of the atom from the centrosymmetric 0.5$a$ position (black diamonds) and the angle between lattice vectors (red squares), (b) reports the changes in the unit cell volume (blue circles) and (c) the corresponding phonon frequencies (green circles). The negative numbers indicate imaginary values of the frequency. Dashed lines denote the phase transition. (d) Electron localization function plotted in the plane of distortion [0-11] for the centrosymmetric (upper left) and distorted (upper right) structures without doping and assuming the same rhombohedral unit cell for both. The bottom panels report $p$-type (left) and $n$-type doped (right) polar phases. In this case, the unit cells correspond to the relaxed volumes displayed in (b). The highest localization is displayed in red, thus the orange/redish lobes are interpreted as the lone pairs of Ge.
    }
\end{figure*}

The idea of realizing spin Hall effect in a device based on FERSC raises the question whether ferroelectricity, the key property of these materials, can be sustained upon doping which is required to achieve a non-zero SHC. In fact, the existence of ferroelectric metals has been debated for several years,\cite{anderson_blount, benedek_review} as not only do the conduction charges screen the external electric fields, but they also suppress the Coulomb interactions that give rise to polar distortions. Although true ferroelectric metals are rare,\cite{shi, iniguez, rondinelli} it has been shown that the \textit{polar phase} can survive in some materials upon doping. For example, ferroelectric distortions in BaTiO$_3$ are sustained up to a critical concentration of 0.11$e$ per unit cell volume (u.c.),\cite{tsymbal_wang, iwazaki_bto} whereas in BiFeO$_3$, PbTiO$_3$ and other lone pair driven ferroelectrics the charge injection even enhances the polar structures.\cite{bifeo3, lone_pair} Since the response to doping will depend, in general, on the origin of ferroelectricity (e.g. short-range vs long-range interactions among other factors),\cite{priya_mahadevan, chinese} a careful analysis is needed in order to confirm the compatibility between SHE and the polar phase.

The ferroelectric displacement in both GeTe and SnTe is driven by the lone pair of the cation. As it has been shown in detail by Waghmare \textit{et al.}, the stereochemical activity of the latter is determined by the interacting $s$ and $p$ states of the cation mediated by the mixing with the anion $p$ states.\cite{spaldin} The strictly chemical origin of the ferroelectricity suggests that the displacement might be sustained even in the presence of free charges screening the Coulomb interaction. We note, however, that the previous studies demonstrating the persistence of polar distortion in lone pair driven ferroelectrics were focused on electron doping, while GeTe and SnTe are always $p$-type doped due to the vacancies that naturally form in these materials.\cite{vacancies} In the analysis below, we consider both signs of charge in order to provide a complete picture of the physics behind.

Figure \ref{doping}(a) shows the evolution of polar displacement in GeTe as a function of charge carrier concentration. We can immediately notice that $n$- and $p$-type doping yield two completely different structural scenarios. In case of the former, the displacements are indeed enhanced, similarly as in the previous studies of lone pair driven ferroelectrics. In contrast, $p$-type doped structures reveal a rather fast decay of polar distortions which vanish above the critical hole concentration of $n=0.13|e|$/u.c. corresponding to about $2.2\times10^{21}$/cm$^{3}$, closely resembling the behavior of BaTiO$_3$ with ferroelectric instability ensured by the Coulomb force.\cite{tsymbal_wang} This indicates that the lone pairs in GeTe might become less active upon $p$-type doping. We also notice that the volumes of the unit cell, Fig.\ref{doping}(b), follow quite similar trend, i.e. a larger displacement tends to increase the volume, while paraelectric structures favor smaller unit cells. Finally, ferroelectric to paraelectric transition in $p$-type doped regime is further confirmed by the phonon calculations of the doped cubic cells shown in Fig.\ref{doping}(c). Ferroelectric instabilities manifest as the imaginary frequencies of phonons that soften close to the point of phase transition.

In order to shed more light on these intriguing findings, we have calculated the electron localization function (ELF) which helps to identify regions associated with electron pairs. Figure \ref{doping}(d) compares ELF plotted for selected configurations from Fig.\ref{doping}(a-c) as well as for an undistorted structure. The lone pairs are seen as pronounced red regions around the Ge atoms, rather spherical in the centrosymmetric configuration (upper left) and distorted in the ferroelectric structure (upper right). The asymmetric lobe confirms the activity of the lone pair in the polar phase and demonstrates dominant role that it plays in the mechanism of ferroelectricity. Furthermore, the comparison of the heavily $p$- and $n$-type 'ferroelectrics' (bottom right and left, respectively) clearly shows that in the former the lobes resemble those observed in the centrosymmetric structure. In contrast, the asymmetric lobe in the electron-doped configuration is even more enhanced with respect to the undoped ferroelectric, which indicates a stronger stereochemical activity of Ge.

\begin{figure}[h!]
    \includegraphics[width=\columnwidth]{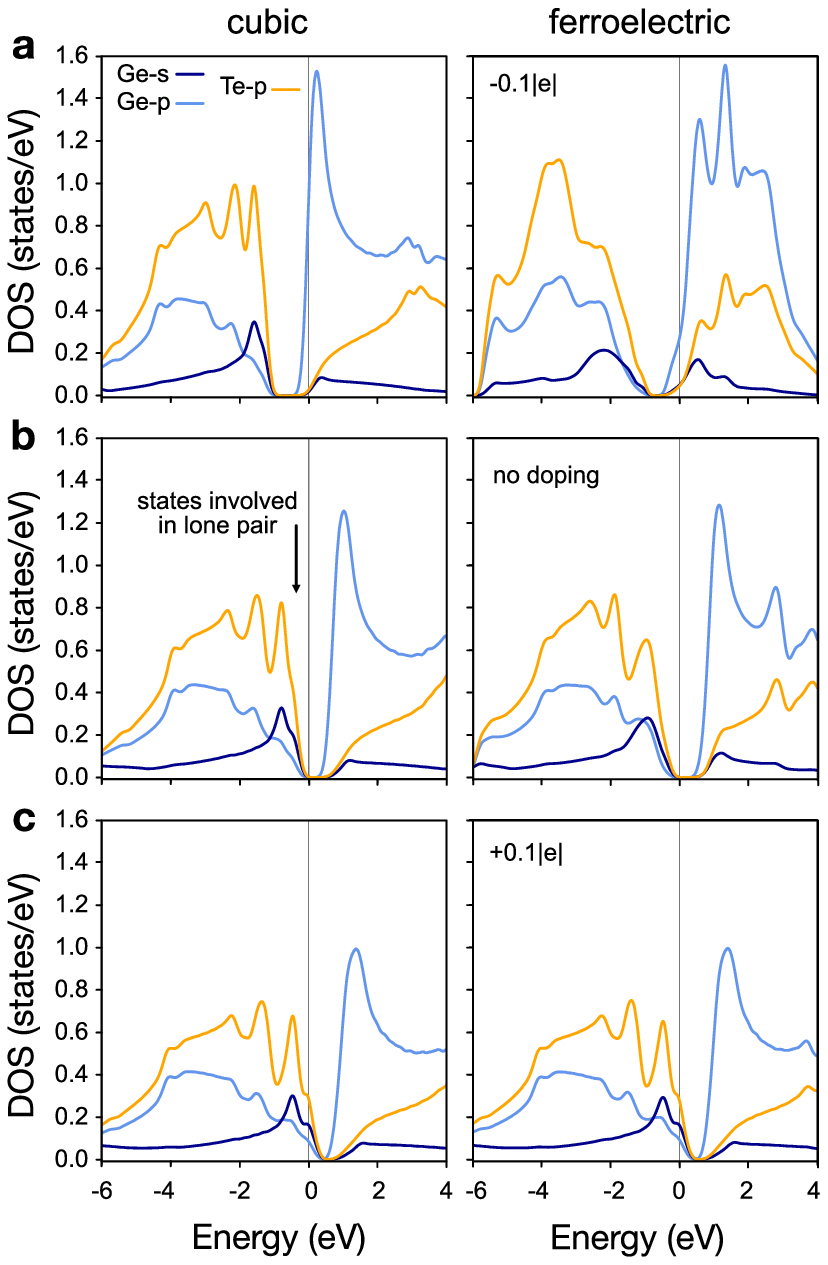}
    \caption{\label{pdos}
  Changes in density of states in doped GeTe. (a) Density of states projected on $s$ and $p$ states of cation (dark and light blue, respectively) and $p$ states of anion (orange) in $n$-type doped cubic and distorted rhombohedral unit cell of GeTe. (b) Same as (a) without doping. (c) Same as (a) with opposite $p$-type doping of 0.1|e|/u.c. All the structures were fully relaxed.
    }
\end{figure}

The reason of the distinct evolution of the hole and electron doped structures can be further rationalized by the analysis of the projected density of states (PDOS) calculated for the corresponding cubic cells (Fig.\ref{pdos}). As discussed in Ref. \onlinecite{spaldin}, the primary cause of the distortion is mixing between $s$ and $p$ states of the cation (dark and light blue, respectively) resulting from a delicate balance of their bonding/antibonding with the anion $p$ states (orange lines). Since this effect occurs predominantly at the top of valence bands, it is thus not surprising that the lone pair remains unaffected by $n$-type doping which shifts the Fermi level in opposite direction (Fig.\ref{pdos} (a)). On the other hand, in the presence of holes as shown in Fig.\ref{pdos} (c), all the involved states reconstruct, slightly hybridizing with conduction bands which in turn suppresses the distortion.

\begin{figure}[h!]
    \includegraphics[scale=1.0]{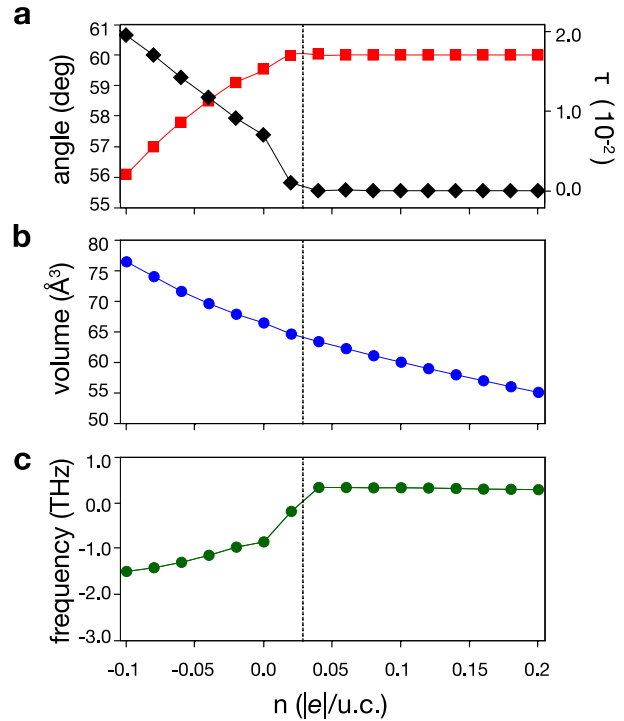}
    \caption{\label{doping_snte}
  Same as Fig. \ref{doping} (a)-(c) calculated for SnTe.
    }
\end{figure}

In SnTe, the calculations of structural parameters with varying carrier concentration revealed trends very similar to those observed in GeTe. We emphasize that the polar displacement, even without doping, is very tiny ($\tau\sim$0.01). Moreover, as can be observed in Fig.\ref{doping_snte} (a), $\tau$ decays much faster than in GeTe; the polar phase is completely suppressed at the carrier concentration of $n=0.03|e|$/u.c. equivalent to around $4.0\times10^{20}$/cm$^{3}$, a doping commonly observed in experiments. We believe that  the  weaker ferroelectricity in this material can be assigned to a smaller and less active lone pair, resulting in a decreased tendency to a covalent bond formation.

Finally, let us comment on the practical realization of polarization reversal which can be challenging in a doped semiconductor. 
Recent experimental study by Kolobov \textit{et al.} based on piezoresponse force microscopy have demonstrated that the ferroelectric switching is possible in $p$-type doped epitaxial GeTe films.\cite{switching_gete} As estimated above, the polar displacements should survive even at the high doping level of over $10^{21}e$/cm$^{3}$ above the typically measured values,\cite{mauri, doping_value, singh} but the applied electric field might still be screened by free carriers induced by the vacancies. We believe that the successful realization of ferroelectric switching in GeTe samples could be related to a strong electric field applied \textit{locally} with the PFM tip which allows to switch the domains. As pointed out by Kolobov \textit{et al.}, the reduced thickness of the film might also play a  role in the overall mechanism of the switching. These indicate that the ferroelectric switching in this material has to be further explored from both fundamental and practical side; ideally, the polarization reversal should be achieved by applying an opposite voltage with the electrode contacts. On the other hand, the ferroelectric switching has never been reported in a similar rhombohedral SnTe, most likely due to the weak ferroelectricity and low temperature of phase transition, which also impedes the realization of FERSC in this material.

\section{Spin Hall effect in doped FERSC}

\begin{figure}[h!]
    \includegraphics[width=\columnwidth]{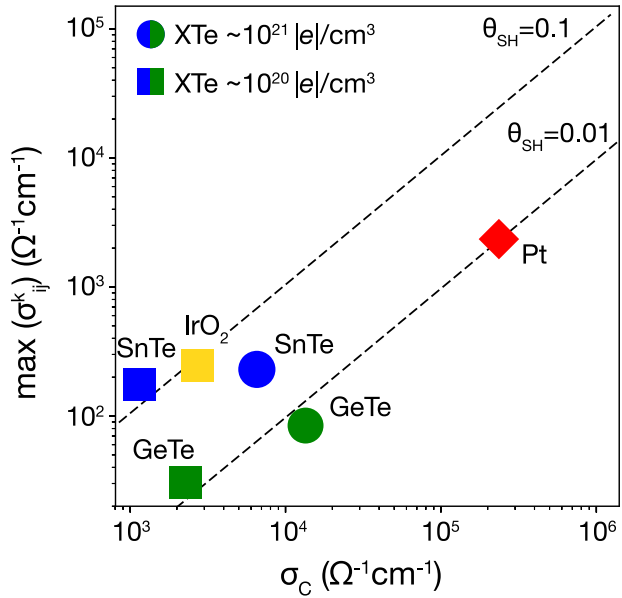}
    \caption{\label{she}
    Spin Hall angles of doped XTe and semimetallic/metallic materials. Spin Hall conductivities were calculated using \textsc{Paoflow} code, whereas charge conductivities $\sigma_{C}$ for specific carrier concentrations were estimated from the experimental data in Ref. \onlinecite{gete_conductivity} for GeTe, Ref. \onlinecite{mandale} for SnTe and Ref. \onlinecite{fujiwara} for IrO$_2$. We note that the actual $\sigma_{C}$ will strongly depend on sample preparation and structure. The dashed lines indicate large (0.1) and moderate (0.01) spin Hall angles.
    }
\end{figure}

Finally, we will focus on the spin Hall effect in the doped structures discussed in the previous section. Figure \ref{she} reports the values of spin Hall angle, defined as a ratio of spin and charge conductivity. We have compared the values estimated for XTe at moderate and high concentrations of holes with the calculations for known spin Hall materials. The highest spin Hall conductivities in GeTe within the considered range of carrier concentrations are approximately 30-80 \unit\, with the maximum value found at $n=0.1|e|$/u.c.=$2.0\times10^{21}|e|$/cm$^{3}$ for $\sigma^{x}_{yz}$. The calculated spin Hall angle $\theta_{SH}\sim$0.01 (green circle) is lower  than in case of Pt and much below the value we estimated for (nodal line) semimetal IrO$_{2}$.\cite{iro2, felser} It can be assigned to the large charge conductivity in doped GeTe. On the other hand, $\theta_{SH}$ is higher at $n=0.02|e|$/u.c. $\approx 3\times10^{20}$/cm$^{3}$ (green square), as the charge conductivity changes faster than the spin Hall conductivity. This confirms that the sizable spin Hall effect could be explored in GeTe at hole concentrations of $10^{20}-10^{21}/$cm$^{3}$ which genuinely sustains the polar phase and is typically observed in experiments.

Analogous trends can be noticed for SnTe displayed as blue square and circle in Fig.\ref{she}. We have found that the spin Hall conductivities are much above those estimated for GeTe, similarly as in case of previously reported 2D-XTe.\cite{2dmaterials} Notably, the values of SHC continuously grow with doping, achieving 170-250 \unit\, for hole concentrations of $0.02-0.1|e|/$u.c. which results in very high spin Hall angles of 0.03-0.15. However, the polar distortions of SnTe are suppressed already at $0.02|e|/$u.c., thus the giant spin Hall effect should rather be attributed to the paraelectric phase. In fact, the estimated spin Hall angle is in good agreement with the experimental value of 0.01 measured in the high-temperature rocksalt structure by Ohya \textit{et al.}\cite{ohya_snte} Although it has been suggested that the topological surface states could enhance the efficiency of spin Hall effect, our analysis shows that the large spin Hall angle can be entirely assigned to the strong SOC of bulk electronic states in the cubic structure. Thus, despite the realization of FERSC seem difficult in this material, it is extremely promising for spintronics applications based on spin Hall effect.

\section{Conclusions}
In summary, we have studied spin Hall effect in prototype Rashba ferroelectrics GeTe and SnTe. Although such an idea might appear contradictory, we have demonstrated that the spin Hall effect could be indeed observed in a doped polar (or ferroelectric) material. We have explored several  aspects related to the realization of SHE in polar structures. First, we have noticed that the Rashba splittings of the electronic bands can enhance spin Hall conductivity, but they play a minor role at the hole concentrations usually observed in the considered materials. Second, we have investigated the evolution of polar distortions under carrier doping and found that the lone pair driven ferroelectricity is robust against electron injection, whereas in the presence of holes the polar displacements can be sustained up to approximately 0.13 (0.03) $|e|$/u.c. in GeTe (SnTe). Finally, we have reported spin Hall angles for both materials at the typical concentrations of holes $10^{20}-10^{21}$/cm$^{3}$. The values in GeTe are close to 0.01 and refer to the polar phase, while giant spin Hall effect in SnTe characterized by the angles of $\sim$ 0.1 is likely to be found in paraelectric structures. In conclusion, our results indicate that the spin Hall effect could be realized in polar material, which can result in development of novel spintronics devices based on the interplay of efficient charge-to-spin conversion, Rashba-related physics and ferroelectricity. We also believe that our study will stimulate further search of FERSC with complementary spintronics properties.

\section*{Acknowledgments}
We are grateful to Naiwrit Karmodak, Christian Rinaldi, Riccardo Bertacco, Marcio Costa and Frank Cerasoli for valuable discussions. The members of the AFLOW Consortium  (http://www.aflow.org) acknowledge the grant ONR-MURI N000141310635. Finally, we acknowledge the High Performance Computing Center at the University of North Texas and the Texas Advanced Computing Center at the University of Texas, Austin.

\section*{Author contributions}
J.S. coordinated the project, providing calculations, data analysis and writing the manuscript; H.W. initiated the project and performed calculations; P.G. performed calculations; M.B.N. and S.C. analyzed the results; S.P. and M.B.N. conceived the idea of spin Hall effect in Rashba ferroelectrics. All the authors contributed to the discussion of the results.

\section*{Data availability}
Data are available from the authors upon reasonable request.

\section*{Code availability}
The \textsc{paoflow} code can be downloaded from [http://aflowlib.org/src/paoflow/].


%
\end{document}